\documentclass[twocolumn,aps,floatfix]{revtex4}

\usepackage{psfig}
\usepackage{epsfig}
\usepackage{natbib}

\begin{document}

\title{Averages and critical exponents in type-III intermittent chaos}
\author{Hugo L. D. de S. Cavalcante} 
\author{J. R. Rios Leite}
\affiliation{Departamento de F\'{\i}sica,  Universidade Federal de Pernambuco, 
50670-901 Recife, PE Brazil}
\date{\today}

\begin{abstract}

 The natural measure in a map with type-III intermittent chaos is used to 
define critical exponents  for  the average of a variable  from a dynamical 
system near bifurcation. Numerical experiments were done with maps and 
verify the analytical predictions. Physical experiments to test the usefulness
of such exponents to characterize the nonlinearity at bifurcations
were done in a driven electronic circuit with diode as
nonlinear element.
Two critical exponents were measured: $\nu= 0.55$
for the critical exponent of the average of the voltage across the diode and 
$\beta =0.62 $ for the exponent of the average length of the laminar phases.
 Both values are consistent with the predictions of a type-III intermittency 
of cubic nonlinearity.
The  averages of variables in intermittent chaotic systems is a technique 
complementary to the measurements of laminar phase histograms, to identify  the
nonlinear mechanisms. The averages exponents may have a broad application in 
ultrafast chaotic phenomena.
\end{abstract}

\maketitle

DOI 10.1103/PhysRevE.66.026210



\vspace{1cm}

\section{Introduction}

Intermittent chaos is the phenomenon shown by systems exhibiting long sequences
of periodic-like behaviour, the laminar phases, separated by comparatively
short chaotic eruptions. 
Intermittent chaotic systems have been extensively studied since the
original proposals by Pomeau and Manneville
\cite{pomeau80,berge84,manneville90,kim98} classifying type-I, II, and III
instabilities when the Floquet multipliers of a map
crosses the unity circle. 
Although other mechanisms may occur leading to intermittency 
 these three cases are the most simple and most frequently encountered
in low dimensional systems.
The class of  intermittent chaos studied by Pomeau and Manneville includes
the tangent bifurcation, leading to intermittency of type I, when the
Floquet's Multiplier for the associated map crosses the circle of complex
numbers with unitary norm through $+1$; the Hopf bifurcation, leading to
type-II intermittency, which
appears as two complex eigenvalues of the Floquet's Matrix cross the unitary
circle off the real axis; and the sub-critical period doubling, leading to
type-III intermittency, whose critical Floquet's Multiplier is $-1$.

Many experimental evidences for these
intermittent behavior have appeared in the literature. The type-III 
 has been reported for
electronic nonlinear devices
\cite{jeffries82,fukushima88,ono95}, lasers \cite{tang91}, and
biological tissues \cite{griffith97}. A signature for this
intermittency is
given by the critical exponent describing the dependence of
average length of nearly periodic phases, that is, laminar phases, with
the control parameter. Histograms of number of laminar phases longer than
 a given duration are related to the normal form nonlinearity describing
the system \cite{berge84,kodama91}. Different nonlinear power in the model
imply different exponents, as described by Kodama
{\it et al.}~\cite{kodama91}. 
Kim {\it et al.}~\cite{kim98} studied the exponent for the average length of
laminar phases as function of the reinjection probability.
Herein another average is explored to
characterize the intermittency and its nonlinearity. It is the average of
one variable of the system. Near bifurcation, approximate expressions
for the natural measure, or probability density, are obtained
 for a map with type-III intermittency, in section \ref{sec:map}.
Critical exponents are defined from analytical approximations
and shown to be useful in
characterizing the bifurcation.

Numerical experiments with the maps are presented 
in section \ref{sec:numerical},
verifying the proposed exponents.
To test in a real physical system,  a nonlinear circuit with a diode,
similar to the one used in the early demonstrations of chaotic universal
properties \cite{jeffries82,linsay81,testa82,buskirk85}, was set up as 
described in section \ref{sec:circuit} and
used to verify the  exponents. The  voltage
 across the diode,  which is a dynamical variable of the system,
was simultaneously measured in time series  and average.
The  type-III intermittency bifurcation is well
characterized both,
using the experimental  next peak value map and the average.

\section{Normal Form map with type-III intermittency}
\label{sec:map} 

To establish the new critical exponents for the averages one begins
with the normal form of the map that has type-III intermittency
\cite{berge84}

\begin{equation}
M(X)= -(1+\epsilon^\prime)X + \alpha X^2 +\eta^\prime X^3
\label{eq:map1}
\end{equation}
The bifurcation, when $X=0$ ceases to be a stable fixed point,
  occurs at $\epsilon^\prime=0$. The second application of this map,
  in the approximation of small $\epsilon^\prime$ and X is given by

\begin{equation}
M^2(X)= (1+2\epsilon^\prime)X + \eta X^3
\label{eq:map2}
\end{equation}
where $\eta=-2(\eta^\prime + \alpha^2)$.
If this coefficient is positive and $\epsilon^\prime > 0$
one has type-III intermittency around $X=0$,
provided a reinjection mechanism is introduced in
eqs.~(\ref{eq:map1})~and~(\ref{eq:map2}).
When $\alpha \neq 0$ the map of eq.~(\ref{eq:map1}) is nonsymmetrical in
$X$.
 Generally, any odd exponent in the nonlinear term of
eq.~(\ref{eq:map2}) leads to the intermittency. 
 Thus, a map (second iterate) defined in the interval $[0,1]$ with the 
type-III intermittency is \cite{manneville90}

\begin{equation}
X_{n+1}=\left[ (1+\epsilon)X_n + X_n^z\right]\ \, \mathrm{mod}\ 1
\label{eq:map3}
\end{equation}

\begin{figure} [htbp]
\centerline{\psfig{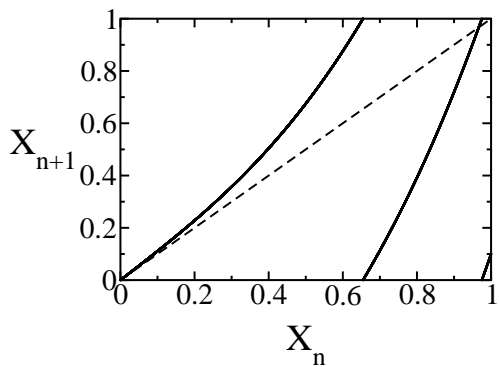}}
\caption
{Map with the  normal form  that gives  type-III intermittency.
The value $ z=3$ was used in eq.~(\ref{eq:map3}). }
\label{fig:map3}
\end{figure}

The value of $z \ge 3$ describes the nonlinear dependence of the subcritical
bifurcation at $\epsilon = 0$.
For $\epsilon \ll 1$ the fixed point $X=0$ is unstable, but many iterates
of the map, shown in Figure~\ref{fig:map3}, falls near zero.
These are called the laminar phase iterates.

  For small $\epsilon$ the natural measure  is
obtained, following the steps of Manneville \cite{manneville90a}, as

\begin{equation}
\mu(\epsilon, z, X, dX)=
\left[\int_0^1{\frac{dX}{\epsilon+X^{(z-1)}}}\right]^{-1}
{\frac{dX}{\epsilon+X^{(z-1)}}}
\label{eq:measure}
\end{equation}
Using this measure,   the average of $X$ is

 \begin{equation}
\left< X(\epsilon, z) \right> =
\left[\int_0^1{\frac{dX}{\epsilon+X^{(z-1)}}}\right]^{-1}
{\int_0^1\frac{X\ dX}{\epsilon+X^{(z-1)}}}
\label{eq:average}
\end{equation}
Its dependence on $\epsilon$ and $z$ can be
analytically established for small $\epsilon$.

 The simple case $z=3$ reduces to
\begin{equation}
\left< X(\epsilon, z=3) \right> \approx
- \frac{1}{\pi}\epsilon^{1/2}\ln \epsilon
\label{eq:aveXz3}
\end{equation}

For $z>3$ the general asymptotic expression is
\begin{equation}
\left< X(\epsilon, z) \right> \propto
\epsilon^{\nu}
\label{eq:aveXz}
\end{equation}
where $\nu$ is the new exponent, whose value is

\begin{equation}
 \nu=1/(z-1)
\label{eq:exp}
\end{equation}
This is the exponent that can be extracted from numerical and
experimental systems to obtain the value of $z$.
Similar expressions can be obtained for the second iterated map with
negative values of $X$. If the original first iterated map, giving
eq.~(\ref{eq:map3}) is fully odd in $X$ the total average must be zero.
Once the reinjection or the map is not symmetrical, i.~e., the value of
some even nonlinear power (less than $z$) is non zero,
the exponents of the averages can be obtained from the second iterate maps
as eq.~(\ref{eq:map3}). A detailed study of these symmetry properties
will be published elsewhere \cite{hugo2001}.
One must recall that, in type-III intermittent chaos, an exponent already
exists in the literature  for the average length of laminar phases,
 given by \cite{kodama91,ono95}  $\beta=(z-2)/(z-1)$.
 The simple relation exists,
 giving $\nu + \beta=1$. Therefore, as $z$ varies through $3, 5, \ldots$
 the relative variation of these exponents are
 $\Delta \nu / \nu > \Delta \beta / \beta$. 
This means equal or better sensitivity
 in the determination of $\nu$ with respect to $\beta$. 
The most important fact is that
 those two exponents should be searched for independently in any experiment.

\section{Numerical Averages and exponents in the map}
\label{sec:numerical}

\begin{figure} [htbp]
\centerline{\psfig{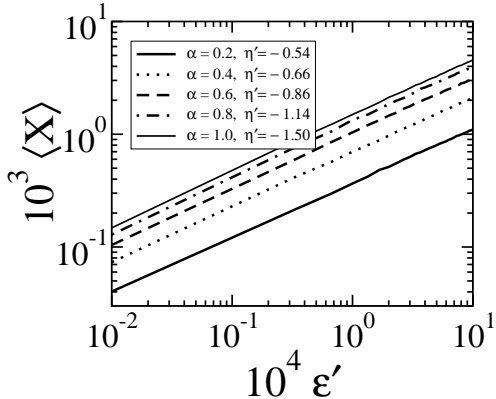}}
\caption
{ Average of $X$ from the map in equation~(\ref{eq:map1}), with the
modulo $1$ reinjection and calculated with
$10^6$ iterates at each of $10^3$ steps of the control parameter
$\epsilon$. The log-log plot shows the
same slope for all averages, independent of the value
of $\alpha$ and $\eta^\prime$. }
\label{fig:avealpha}
\end{figure}

To test the predictions of eq.~(\ref{eq:exp}) numerically the map of
eq.~(\ref{eq:map1}) was restricted to $|X| <1$ by a modulo one
reinjection: When the iterate gives $X>1$ or $X<-1$ the
integer part is subtracted.  Averages were obtained directly from
iterates of the map starting from random initial condition. A transient
of $2\times 10^4$ iterates was eliminated from a total of $10^6$
calculated values at each of  $10^3$ steps of $\epsilon^\prime$, between
zero and $10^{-3}$.  For the average to be non null the value of $\alpha$
has to be different of zero.  This breaks the symmetry between positive and
negative values of X.  Once the average is non null, its behavior for
small $\epsilon$ follows  eq.~(\ref{eq:aveXz3}). This is shown in
Figure~\ref{fig:avealpha}.
 The $\alpha =0$ case, where the average is always zero, is not
represented.
  The same slope is
obtained for the averages with different values of $\alpha$ and
$\eta^\prime$. The resulting averages have the same dependence in
$\epsilon$ as the one obtained from the map of second iterates
(eq.~(\ref{eq:map3})), calculated with the same number of iterates; all
coinciding with the expression of eq.~(\ref{eq:aveXz3}).

\begin{figure} [htbp]
\centerline{\psfig{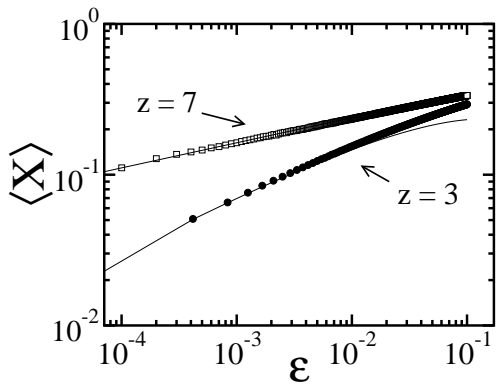}}
\caption
{ Average of $X$ from the map in equation~(\ref{eq:map3}), calculated
with $2\times10^7$ iterates at each of $10^3$ steps of the control
parameter $\epsilon$. 
White squares are used for $z=7$ and filled circles for $z=3$.
The thin lines correspond to the predicted values given by 
eqs.~(\ref{eq:aveXz3}) and (\ref{eq:aveXz}).
} \label{fig:avez}
\end{figure}

Figure~\ref{fig:avez} shows  numerically
calculated
averages obtained directly from  iterates of the map eq.~(\ref{eq:map3}),
when $z=3$ and $z=7$. Again the initial condition at each value of
$\epsilon$ was taken at  random. A transient of $10^5$ iterates
was eliminated from a total of $2 \times 10^7$ calculated values at each
of  $10^3$ steps of $\epsilon$, between zero and $10^{-1}$.

For $\epsilon < 10^{-2}$  the $z=3$ and $7$ results are very well 
superimposed by the analytical curves made from  equation~(\ref{eq:aveXz3}),
 with $\nu=0.50$, and  equation~(\ref{eq:aveXz}), with $\nu=1/6\approx 0.17$,
respectively.  The  two behaviors  are
clearly distinguished in the log-log plot.
 For the purpose of comparison with the already established exponents one
should  notice that the  exponents
for the average laminar phase are given in \cite{kodama91} as $(z-2)/(z-1)$
that is $\beta=0.50$ and $\beta=0.83$, respectively, verifying
$\nu+\beta=1$.

Up to this point all results have concerned discrete maps.
Dynamical fluxes with associated maps having intermittent chaos
show similar critical exponents for the average of its continuous variables.
This has been verified numerically with tangent bifurcation of
type-I intermittency \cite{hugomapaave2000,hugoave2000} and will be
demonstrated here for a physical chaotic oscillator.
For the type-III intermittency the map extracted from a flux has the
form of eq.~(\ref{eq:map3}) for its second iterates.
A discretely sampled continuous flux may be viewed as a finite set of 
discrete stroboscopic maps, all with the same stroboscopic frequency but 
stroboscopic phases ranging from 0 to $2\pi$ in uniformly separated steps. 
The number of such maps gives the ratio of sampling frequency to stroboscopic
frequency. The time average of the continuous flux is the average of the
time averages of these maps.
If a map from this series is nonsymmetrical, only in very special
cases it would happen that this asymmetry  be exactly canceled by the 
summation on the other maps. Thus, in general, its expected that 
 if the flux leads to a nonsymmetrical map (no odd symmetry with respect
to the unstable fixed point) the average of its continuous variable will
exhibit the same exponent $\nu$ of eq.~(\ref{eq:exp}).

\section{Experimental exponents in a nonlinear circuit with intermittency} 
\label{sec:circuit} 

\begin{figure} [htbp]
\centerline{\psfig{figure=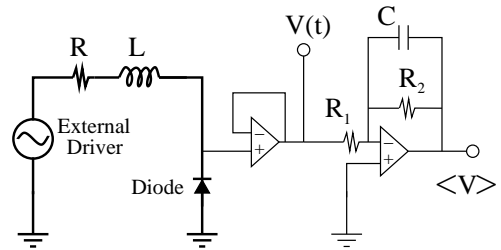,width=6.5cm}}
\caption
{ Experimental diagram of the RLD circuit showing type-III intermittent
chaos.
 The voltage across the diode was the dynamical variable 
measured both in its peak value, to make maps, and integrated 
to give the averages.
The circuit is drawn with thick lines, while thin lines are the high
impedance probe and  integrator.
}
 \label{fig:circuit}
\end{figure}

 The physical experimental  system to test the above exponents
consisted of an RLC series circuit, where a
 nonlinear capacitance was  implemented by a p-n junction diode
\cite{testa82,buskirk85}.
The circuit, presented in figure \ref{fig:circuit}, 
 was driven by a frequency and
amplitude controllable  oscillator with output
impedance of $50$ Ohms. An inductor of $0.1$
Henry, an external  resistor of $13$ Ohms and a (typical 1N4007) diode
formed the  circuit, whose linear oscillation regime had
   resonance frequency at $150$ kHz.

Dynamical bifurcations were produced by scanning the external
drive frequency.
 Time series of the value of the
voltage across the diode and the current in the circuit were collect with
a $12$ bits resolution converter. The sampling rate was $10^7$ sample/s.
Thus $200$ points were  saved on each oscillator cycle.
Capturing $10^4$ points at each value of the control frequency,
a simple software searched for the maxima in the
series.
In wide range scans the  typical bifurcation
diagrams given by the peak value of the voltage across the diode,
show clearly the well known results of period doubling
cascades, chaotic windows and tangent bifurcations from chaos into
periodic windows. \cite{jeffries82,testa82,buskirk85,ono95}.

\begin{figure} [htbp]
\centerline{\psfig{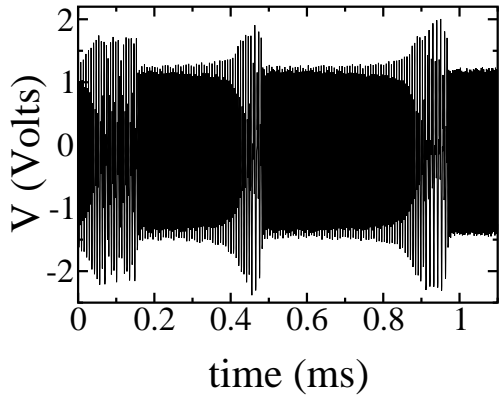}}
\caption
{ Voltage pulses measured across the diode in the RLD circuit.
Successive maxima alternate their position respective to the $1$ V value until
a chaotic burst occurs and reinjects back to the vicinity of the unstable 
orbit, typical of type-III intermittency }
\label{fig:pulses}
 \end{figure}

A specific bifurcation, with intermittency and no bistability,
was found and studied,
scanning the external oscillator frequency   from $48708$ Hz
to $48768$ Hz by  $300$ equal small steps. The
drive voltage amplitude was fixed at $2.7$~V.
The circuit was mounted inside an  isolating box to prevent thermal drift
effects \cite{donoso97}.
 The value of a control parameter $\epsilon$ is
obtained as the frequency detuning step divided by the
critical frequency of the bifurcation, $48762$~Hz.
Thus $\epsilon$ varied in the range of $10^{-4}$--$10^{-3}$.
A segment of the voltage pulses is shown in figure~\ref{fig:pulses}. The
signature of type-III intermittency is observed with the laminar events 
corresponding to the uniform oscillations alternating their peak value around
1V. 

\begin{figure} [htbp]
\centerline{\psfig{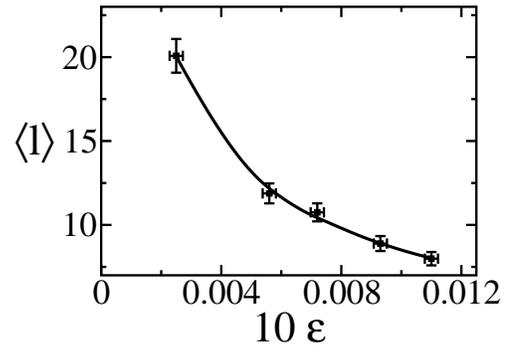}}
\caption
{ Measured average  lengths of laminar oscillations in
 the driven circuit with a diode.
The thick line is the  theoretical  fitting for type-III
intermittency  with exponent $\beta=0.62$ }
\label{fig:histogram}
 \end{figure}

Multibranch maps constructed from the maxima of the voltage across the
diode, extracted from series with $6 \times 10^5$, could be
approximated to one-dimensional ones, expected in the limit of infinite
dissipation. Second return maps and laminar phase histograms, not shown
here, indicate  type-III intermittency. Figure~\ref{fig:histogram} shows
the average length of laminar phases for different values of $\epsilon$.
Each histogram used was obtained with $6 \times 10^5$ points.
A theoretical fitting \cite{kodama91} is best with an exponent
$\beta=0.62$.
Notice that for the
map the predictions are $\beta=0.5$ for $z=3$ and $\beta=0.75$ for $z=5$.
However the experimental data always has excess of laminar events
identified with short length \cite{griffith97} and
this effect  gives a bigger experimental value for $\beta$ in the fittings.
Therefore, $z=3$ is the best odd value.The excess of short laminar 
phase events may be related to the
finite dissipation rate in the phase space of the experimental system and
the consequent non unidimensional maps and also to a nonuniform density
of probability for the reinjection, as discussed by 
Kim~{\it et al.}~\cite{kim98}.

\begin{figure} [htbp]
\centerline{\psfig{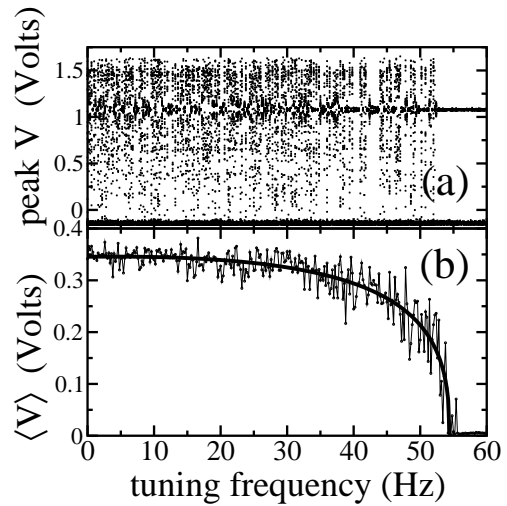}}
\caption
{ (a) Peak voltage across the diode in the type-III bifurcation of the circuit.
(b) Simultaneously measured average voltage.
The thick line is a  fitting of eq.~(\ref{eq:aveXz3}) with exponent
$\nu=0.55$ } \label{fig:ExpAve}
\end{figure}

 The peak value of the voltage, which gave the second return map
and the histograms is
shown in Figure~\ref{fig:ExpAve}(a).
The laminar phases are the oscillations with repetitive visits to the
maximum value near $1V$ in the figure. 
As the segments represented in figure~\ref{fig:pulses}(a) are short in time
for many values of parameters they  show a single laminar event with all points
accumulated around the unstable orbit.
Also shown is the  simultaneously acquired  average of the voltage across 
the diode. It was  obtained  with a simple  electronic integrator,
having a time constant of $3$ seconds.
To account for long laminar phase events and (which is equivalent)
decrease the average fluctuations  the scan lasted  50 minutes.

   The experimental average voltage was
 fitted to the  expression
\begin{equation}
 \left<X(\epsilon)\right> = - {\epsilon}^{\nu}
\ln{\epsilon}
\label{eq:exave}
\end{equation}
The exponent $\nu=0.55$ gave an excellent agreement with
the experimental plot,
as shown in figure~\ref{fig:ExpAve}.
Attempts of  fittings to higher
values of $z$, i.e., to eq.~(\ref{eq:aveXz}), failed.
It is worth noticing that the value predicted for $z=5$ is $\nu=0.25$.
Thus, the experimental average consistently verifies $z=3$ for the
nonlinearity of this bifurcation in the circuit.
The confidence for the experimental values of $\beta$ and $\nu$ from fittings
 using standard $\chi^{2}$ procedure is better than 2\%. Therefore,  the 
result $\beta + \nu = 1.17 \pm 0.02$ shows that a 
discrepancy remains between the experiments and the unidimensional map model.
While the bigger value obtained for $\beta$  has been attributed 
 to an excess of short 
laminar phase events~\cite{kim98} 
no such study of deviations exists for the exponent in the averages.

\section{Conclusion}

In conclusion, critical exponents for the
averages of one dynamical variable are
established analytically  for type-III
intermittent chaotic maps. Those exponents are directly related to the
nonlinear power law of the normal form of the maps.
They have a simple relation with the exponents of
the average length of laminar iterates
in the same systems.
All these properties were verified in  numerical experiments with maps.

Physical experiments with a continuous flux were also done to demonstrate
the exponents in averages.
The average voltage across a diode in a chaotic electronic
circuit was measured while the drive frequency was scanned through
bifurcations. A  bifurcation from chaos into periodic
pulsation,  shown to be type-III intermittent, gives an exponent
in agreement  with the cubic nonlinearity. This result is
consistently verified in the exponents of the average length of laminar
phases, extracted from histograms of the peak pulse voltages in the circuit.

Averages of dynamical variables have been proposed to get the signature
of the Lorenz chaos bifurcation \cite{lawandy87},  have been
numerically studied in critical bifurcations \cite{rajasekar00}, and
experimentally measured in bifurcating pulsed lasers \cite{oliveira96}.
However no systematic study of critical exponents have been done.

The technique of measuring averages of dynamical variable in
intermittent chaos is a complementary
procedure to investigate bifurcations of nonlinear systems.
The experimental average may also be advantageous  
 when detection noise 
for an specific variable has a bandwidth  overlapping the  
frequency bandwidth of the chaotic oscillations.  
For systems with high frequency noise it is naturally bound 
to be more sensitive.
Its experimental motivation is therefore enhanced to characterize
ultrafast chaotic oscillators, like diode lasers, using slow time
 detection techniques.
One extension relevant to the  work presented here is the study of the 
exponents for the averages in bifurcations with nonuniform reinjection
in the intermittency, as studied by Kim {\it et al.}~\cite{kim98}.
 A relation between  exponents of the averages  
and the exponents of the average length of laminar phases events     
should exist generalizing the $\nu + \beta = 1$ given here.
Another potential use for the averages near bifurcation would be 
to complement the confrontation between experimental data and model as 
extracted from nonlinear  data analysis \cite{hegger98,timmer2000}. 
The models inferred from the data analysis must have bifurcations 
consistent with the experimental system. These bifurcations
could be tested comparing both the  exponents of laminar events and the 
exponents of averages of dynamical variables. 
The   Critical exponents have been introduced for many
types of bifurcations in chaotic systems \cite{grebogi86}. Their presence
in simple, experimentally accessible, statistical properties, as the
averages and its higher moments, are under investigation.
The  earliest citation of averages in chaos  can be traced to
 the original propositions
of unpredictability in deterministic chaos \cite{lorenz64}.

\section*{acknowledgments}
 Work partially supported by Brazilian
Agencies: Conselho
Nacional de Pesquisa e Desenvolvimento (CNPq) and
Financiadora de Estudos e
Projetos (FINEP)


\end{document}